\documentclass[a4paper,prl,superscriptaddress,showpacs,twocolumn]{revtex4}
\usepackage[dvips]{graphicx}
\usepackage{amssymb}
\newcommand{\ped}[1]{\ensuremath{_{\rm #1}}}
\newcommand{\apex}[1]{\ensuremath{^{\rm #1}}}

\begin{document}

\title{Evidence for gap anisotropy in CaC$_6$ from directional point-contact spectroscopy}

\author{R.S.~Gonnelli\email{E-mail:renato.gonnelli@polito.it}}
\author{D.~Daghero}
\author{D.~Delaude}
\author{M.~Tortello}
\author{G.A.~Ummarino}%
\affiliation{Dipartimento di Fisica and CNISM, Politecnico di
Torino, 10129 Torino, Italy}
\author{V.A.~Stepanov}
\affiliation{P.N. Lebedev Physical Institute, Russian Academy of
Sciences, 119991 Moscow, Russia}
\author{J.S.~Kim}
\author{R.K.~Kremer}
\affiliation{Max-Planck-Institut f\"{u}r Festk\"{o}rperforschung,
D-70569 Stuttgart, Germany}
\author{A.~Sanna}
\affiliation{ Institut f{\"u}r Theoretische Physik, Freie
Universit{\"a}t Berlin, Arnimallee 14, D-14195 Berlin, Germany}
\author{G.~Profeta}
\affiliation{CNISM - Dipartimento di Fisica, Universit\`a degli
studi dell'Aquila,
 Italy}
\author{S.~Massidda}
\affiliation{SLACS-INFM/CNR and Dipartimento di Fisica, Universit\`a
degli Studi di Cagliari, Italy}

\pacs{74.50.+r, 74.45.+c, 74.70.Ad}

\begin{abstract}
We present the first results of directional point-contact
spectroscopy in high quality CaC$_6$ samples both along the
\emph{ab} plane and in the \emph{c}-axis direction. The
superconducting order parameter $\Delta(0)$, obtained by fitting the
Andreev-reflection (AR) conductance curves at temperatures down to
400 mK with the single-band 3D Blonder-Tinkham-Klapwijk model,
presents two different distributions in the two directions of the
main current injection, peaked at 1.35 and 1.71 meV, respectively.
By ab-initio calculations of the AR conductance spectra, we show
that the experimental results are in good agreement with the recent
predictions of gap anisotropy in CaC$_6$.
\end{abstract}
\maketitle

The discovery of a relatively ``high-$T\ped{c}$'' superconductivity
in graphite intercalated with Ca \cite{Weller,Emery}, Yb
\cite{Weller} and, very recently, Sr \cite{Kim1,Calandra} has
strongly revived the interest in the Graphite Intercalated Compounds
(GICs) and their physics. The Ca-intercalated graphite, CaC$_6$,
with its ``record'' $T\ped{c}$ of about 11.5 K, in particular, has
been the subject of various theoretical and experimental
investigations in the past two years (for a short review of the
initial results see \cite{Mazin1}). One of the most important
questions, however, is still not clear: what is the magnitude and
anisotropy of its superconducting gap? The first experiments (STM,
penetration depth, specific heat) on CaC$_6$ have evidenced a
single, apparently isotropic, s-wave gap with a ratio
$2\Delta/k\ped{B}T\ped{c}$ of the order of the BCS value
\cite{Bergeal,Lamura,Kim2}. Recent tunnel spectroscopy results, on
the other hand, claimed the presence of an isotropic gap with a
magnitude more than 40$\%$ higher than that reported earlier
\cite{Kurter}. The spread of gap values measured up to now range
between 1.6 meV \cite{Bergeal} and 2.3 meV \cite{Kurter}. The
important point is that all these experiments have either probed a
bulk property \cite{Kim2} or a directional one along the
\emph{c}-axis direction \cite{Bergeal,Lamura,Kurter}. As pointed out
in Ref.\onlinecite{Bergeal}, the presence of anisotropic or two-gap
superconductivity in CaC$_6$ cannot be ruled out until tunneling or
point-contact measurements are performed also along the \emph{ab}
direction. On the other hand, recent first-principles density
functional calculations of the superconducting properties of CaC$_6$
have supported the presence of a moderately anisotropic gap which
varies between 1.1 and 2.3 meV, depending on the \textbf{k}-point
and the $\pi$ or interlayer (IL) sheet of the Fermi surface (FS)
involved \cite{Sanna}. Such an anisotropy can be revealed by
directional spectroscopy measurements performed along both \emph{c}
and \emph{ab} direction.

In this paper we present the results of point-contact
Andreev-reflection (PCAR) spectroscopy performed on high-quality
bulk samples of CaC$_6$ \cite{Kim2}. By using a special technique to
realize the contacts, that proved very successful and effective in
the case of MgB$_2$ \cite{Gonnelli1,Gonnelli2}, we were able to
perform directional PCAR spectroscopy at very low temperature both
along the \emph{ab} plane and the \emph{c}-axis direction. Two
different gap distributions in the two directions can reproducibly
be extracted from the experimental data. When compared to the
results of new first-principles calculations these findings
unequivocally prove the anisotropy of the superconducting gap in
CaC$_6$.

The high-quality CaC$_6$ bulk samples used for our measurements were
synthesized by reacting highly oriented pyrolytic graphite (with a
spread of the \emph{c} axis orientation $\leq$ 0.4$\apex{\circ}$)
for several weeks at 350$\apex{\circ}$C with a molten alloy of Li
and Ca \cite{Kim2}. The resulting CaC$_6$ samples have a shiny
golden surface. They are very sensitive to air and moisture which
rapidly damage the sample surfaces. X-ray analysis has shown mainly
the CaC$_6$ reflections with a small ($< 5 \%$) contribution from
impurity phases. Further details on the characterization of the
samples may be found in Ref. \onlinecite{Kim2}. All samples used for
PCAR spectroscopy (size $\approx$ $1\times1\times0.2$ mm$^3$) were
selected to have a very sharp superconducting transition ($\Delta
T\ped{c}(10\%-90\%)=$ 0.1 K) with the onset at $T\ped{c}=$ 11.4 K.

\begin{figure}[t]
\begin{center}
\includegraphics[keepaspectratio, width=1.0\columnwidth]{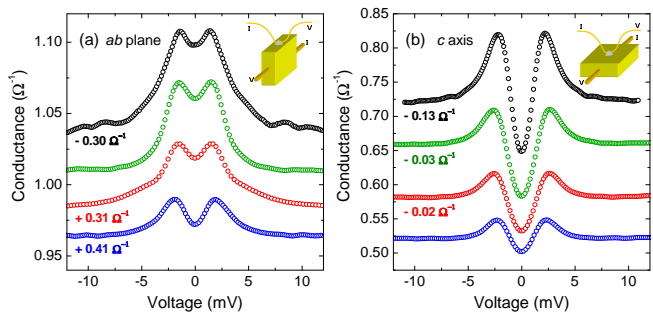}
\end{center}
\vspace{-5mm} \caption{\small{(Color online) (a) Raw point-contact
conductance curves of various \emph{ab}-plane contacts at 4.2 K. For
clarity the curves are vertically shifted of the amount shown close
to each curve. (b) The same as in (a) but for various contacts with
current injection mainly along the \emph{c} axis. In each panel a
sketch of the contact geometry is also shown.}}
\label{Fig:CaC6_fig1}
\vspace{-4mm}
\end{figure}

The point contacts were made by using a non-conventional technique
we called ``soft'' PCAR spectroscopy \cite{Gonnelli1,Gonnelli2}.
Instead of using the standard metallic tip, a very small
($\varnothing\simeq 50$~$\mu$m) drop of Ag conductive paint, put on
the etched or freshly cleaved surfaces of the sample is used as a
counterelectrode. Such contacts are particularly stable both in time
and towards temperature variations and they allow to inject the
current mainly perpendicular to the contact plane. A fine tuning of
the junction characteristics at low temperature can be done by
applying short voltage or current pulses. Further details on the
technique can be found in Refs. \onlinecite{Gonnelli1,Daghero}. Due
to the mentioned high sensitivity of CaC$_6$ samples' surface to
air, the room-temperature preparation of the contact was done inside
a sealed glove bag filled with pure He gas or in a glove box with Ar
atmosphere. After the contact was made the junction was very rapidly
transferred to the cryostat in a sealed container. Contacts were
made either on the flat \emph{ab}-plane surface or on the narrow
lateral side of the samples. Referring to the main direction of
current injection, we call them \emph{c}-axis and \emph{ab}-plane
contacts, respectively (see insets of Fig. 1).

The conductance curves, d$I$/d$V$ vs. $V$, were obtained by
numerical differentiation of the measured $I-V$ curves and
subsequently normalized by dividing them by the normal-state
conductance measured at $T \geq T\ped{c}$. For this reason, in all
the contacts, we therefore carefully studied the temperature
dependence of the conductance in order to determine the critical
temperature of the junction, i.e. the `Andreev critical
temperature', $T\ped{c}\apex{A}$. In an overall of 35 different
contacts, $T\ped{c}\apex{A}$ was found to be $11.3 \pm 0.1$ K, in
best agreement with the bulk $T\ped{c}$'s of the samples and in
contrast with a previous report \cite{Bergeal}. This fact proves the
high quality of samples and surfaces in the contact region. For
simplicity, we will therefore refer to the critical temperatures of
the contacts as $T\ped{c}$ in the following.

Fig.~\ref{Fig:CaC6_fig1} shows several raw conductance curves as
function of bias voltage measured both in \emph{ab}-plane contacts
(a) and in \emph{c}-axis ones (b) at 4.2 K.  The curves show clear
Andreev-reflection (AR) features, an almost flat conductance (at $V
> 8-10$ meV) and no dips that usually are a sign of the failure in
reaching the conditions for pure ballistic conduction in the contact
\cite{Pratap,Naidyuk}. The normal resistance $R\ped{N}$ of all the
good contacts is between 0.75 and 6.4 $\Omega$. By knowing the mean
free paths and the residual resistivities of CaC$\ped{6}$ along the
\emph{ab} plane and in the \emph{c}-axis direction, i.e.
$\ell\ped{ab}=74$ nm, $\ell\ped{c}=4.7$ nm, $\rho\ped{0,ab}=0.8$
$\mu\Omega\cdot$cm and $\rho\ped{0,c}=24$ $\mu\Omega\cdot$cm
\cite{Kim3,Gauzzi} we can apply the Sharvin formula for the contact
resistance in the ballistic regime in order to determine the contact
radius $a=(4 \rho\ped{0}\ell / 3 \pi R\ped{N})\apex{0.5}$
\cite{Naidyuk}. The condition for full ballistic transport ($a \ll
\ell$) is totally verified in \emph{ab}-plane contacts, where
$a\ped{ab} \approx 6-18$ nm. In \emph{c}-axis junctions, where
$a\ped{c} \approx 14-24$ nm but the conductance curves do not show
any sign of heating \cite{Daghero}, the presence of at least 30
parallel contacts in the junction area is expected.

After normalization, the conductance curves were fitted to the
modified 3D Blonder-Tinkham-Klapwijk (BTK) model
\cite{BTK,Plecenik,Tanaka}. In the single-band form it contains
three fitting parameters: The gap $\Delta$, the barrier-height
parameter $Z$ and the broadening $\Gamma$ which accounts for both
intrinsic (quasiparticle lifetime) and extrinsic phenomena that
broaden the AR conductance \cite{Plecenik}.

In order to increase the experimental resolution of our measurements
we decided to perform part of the PCAR experiments at very low
temperature in a Quantum Design measurement system (PPMS) with
$^3$He insert.

\begin{figure*}[t]
\begin{center}
\includegraphics[keepaspectratio, width=0.9\textwidth]{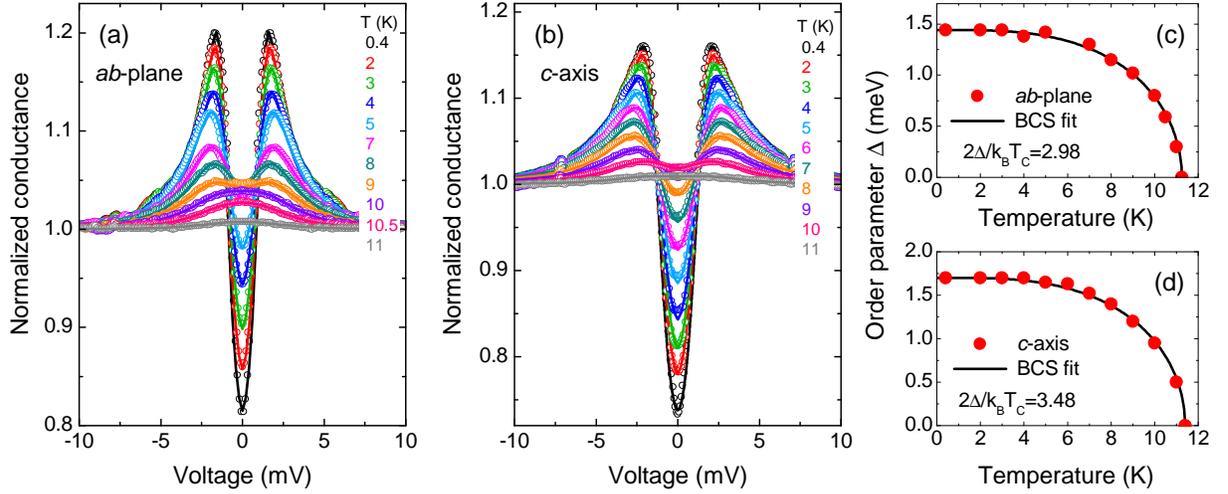}
\end{center}
\vspace{-5mm} \caption{\small{(Color online) Normalized $dI/dV$ vs.
$V$ curves at different temperatures down to 400 mK in an
\emph{ab}-plane contact (a) and in a \emph{c}-axis one (b) (open
circles). Solid lines: best-fit curves given by the single-band 3D
BTK model. Panels (c) and (d) show the temperature dependency of the
order parameter $\Delta$ (full circles) in the \emph{ab}-plane
direction and in the \emph{c}-axis one, respectively, as determined
from the BTK fits shown in (a) and (b). Solid lines are the BCS-like
fits.}} \label{Fig:CaC6_fig2}
\vspace{-4mm}
\end{figure*}

Fig.~\ref{Fig:CaC6_fig2} (a) shows the normalized conductance curves
(circles) of a typical \emph{ab}-plane contact at various
temperatures from 400 mK up to $T\ped{c}$. At any temperature the
single-band 3D BTK model fits the data very well (solid lines).  At
the lowest $T$, the values of the fitting parameters are: $\Delta=$
1.44 meV, $\Gamma=$ 0.61 meV and $Z=$ 0.75. In panel (c) we display
the order parameter $\Delta$ obtained from the data given in (a).
Its temperature dependence almost perfectly follows the BCS-like
expression (solid line) with $2\Delta(0)/k\ped{B}T\ped{c}=$ 2.98
which is sensibly smaller than expected from BCS theory.

In Fig.~\ref{Fig:CaC6_fig2} (b) and (d) we report the same data for
a \emph{c}-axis contact. As for the \emph{ab}-plane case, the curves
are well fitted by the single-band 3D BTK model which gives at 400
mK: $\Delta=$ 1.7 meV, $\Gamma=$ 0.84 meV and $Z=$ 0.97. The
temperature dependence of $\Delta$ is very close to the expected BCS
one with a ratio $2\Delta(0)/k\ped{B}T\ped{c}=$ 3.48, in best
agreement with the weak-coupling BCS value.

It is worth noticing that the $Z$ values observed in \emph{c}-axis
contacts (between 0.74 and 1.01) are systematically greater than
those of \emph{ab}-plane contacts (between 0.48 and 0.75). According
to the 3D BTK model \cite{Tanaka}, this difference can be explained
by the different Fermi velocities of CaC$_6$ in the \emph{ab} plane
($v\ped{ab}=0.54 \times 10^6$ m/s) and along \emph{c} axis
($v\ped{c}=0.29 \times 10^6$ m/s), thus confirming the
directionality of our point contacts.

The AR curves shown in Fig.~\ref{Fig:CaC6_fig1} and
\ref{Fig:CaC6_fig2} are rather small in amplitude, as already
observed in all the ``soft'' PCAR measurements on MgB$\ped{2}$ and
related compounds \cite{Gonnelli1,Gonnelli2,Daghero}, resulting in
$\Gamma$ values substantially greater than those expected for the
quasiparticle lifetime. As recently observed in lithographically
fabricated Cu-Pt-Pb nanocontacts \cite{Chalsani}, this additional
broadening can be explained by the presence of pair-breaking effects
induced by the scattering in a thin disordered layer present at the
NS interface. This is the case of our point contacts, due to a
disordered layer on the surface of Ag grains that also makes the
residual resistivity of the paint be five orders of magnitude
greater than in pure Ag.

\begin{figure}[b]
\begin{center}
\includegraphics[keepaspectratio, width=0.9\columnwidth]{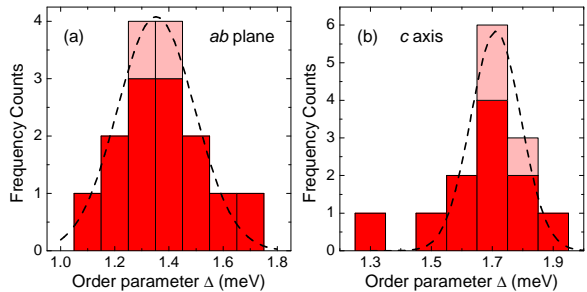}
\end{center}
\vspace{-5mm} \caption{\small{(Color online) Distributions of the
different $\Delta(0)$ values measured in \emph{ab}-plane contacts
(a) and in \emph{c}-axis ones (b) at 4.2 K (red) and at 400 mK
(light red). Dash black lines are the fits of the total distribution
to a Gaussian curve.}} \label{Fig:CaC6_fig3} \vspace{-4mm}
\end{figure}

The reproducibility of the PCAR data was very good. Most of the
contacts, obtained both in $^4$He and in $^3$He cryostat, show
$dI/dV$ curves and temperature dependencies of quality similar to
that presented in Fig.~\ref{Fig:CaC6_fig2}. In 15 \emph{ab}-plane
contacts the order parameter $\Delta(0)$ ranged between 1.1 meV and
1.7 meV with the distribution shown in Fig.~\ref{Fig:CaC6_fig3} (a).
In 14 \emph{c}-axis contacts $\Delta(0)$ ranged between 1.3 meV and
1.94 meV with the distribution shown in Fig.~\ref{Fig:CaC6_fig3}
(b). The figure also shows the Gaussian curves that best fit the
distributions. They are peaked at $\Delta\ped{ab}(0)$=1.35 meV and
$\Delta\ped{c}(0)$=1.71 meV and show standard deviations
$s\ped{ab}$=0.14 meV and $s\ped{c}$=0.08 meV, respectively. The
results in \emph{c}-axis direction are in very good agreement with
the gap values previously reported in
Ref.~\onlinecite{Bergeal,Lamura}. A minority of contacts (3 in
\emph{ab}-plane and 3 in \emph{c}-axis direction) have shown gap
values between 2.1 and 2.4 meV, similarly to the results of
Ref.~\onlinecite{Kurter}.

The complex microscopical nature of our point contacts leaves some
uncertainty about the true direction of current injection,
particularly in the case of contacts on the side faces of the sample
(i.e. \emph{ab}-plane contacts) where, due to the intrinsic
inhomogeneity of the cleaved surface, current injection along
\emph{c} axis is also possible. However, the clear difference
observed between the most probable $\Delta(0)$ values in
\emph{ab}-plane and \emph{c}-axis contacts provides strong evidence
for a gap anisotropy in CaC$_6$.

In order to compare our results with the theoretical predictions of
gap anisotropy in CaC$_6$ \cite{Sanna} we calculated the
Andreev-reflection conductance curves by first-principles methods.
We have a SN junction where S = CaC$_6$ and N = Ag. Let's label with
the suffix $i =$ 1,2,3 the three sheets of the CaC$_6$ Fermi surface
(FS) ($\pi$ and interlayer (IL) bands). If $\mathbf{n}$ is the
unitary vector in the direction of the injected current,
$v\ped{i{\mathbf k},n}= \mathbf{v\ped{i{\bf k}}} \cdot \bf{n}$ are
the corresponding components of the Fermi velocities in the
superconductor at wave vector ${\bf k}$ for band $i$-th. Taking into
account that Ag has a quasi-spherical FS and an almost constant
Fermi velocity $v\ped{N} \neq v\ped{i{\bf k}}$, the corresponding
quantity in the normal metal will be $v\ped{N,n}=v\ped{N}$.
Following Refs.~\onlinecite{Mazin3,Brinkman} we finally obtain the
total AR conductance as:
\begin{equation}
\sigma(E,n)=\frac{\sum\ped{i}\langle\sigma\ped{i{\bf
k}n}(E)\frac{v\ped{i{\bf k},n}\apex{2}} {v\ped{i{\bf k}} [
v\ped{i{\bf k},n} +v\ped{N}]\apex{2}}\rangle\ped{FSi}} {\sum\ped{i}
\langle\frac{v\ped{i{\bf k},n} \apex{2}}{v\ped{i{\bf k}}[v\ped{i{\bf
k},n}+v\ped{N}]\apex{2}}\rangle\ped{FSi}}
\end{equation}
where: $\langle\rangle\ped{FSi}$ is the integral over the $i$-th FS,
i.e.\\

$ \langle\frac{v\ped{i{\bf k},n}  \apex{2}}{v\ped{i{\bf
k}}[v\ped{i{\bf k},n}+v\ped{N}]\apex{2}}\rangle\ped{FSi} =
\int\ped{v\ped{i{\bf k},n}>0} \frac{v\ped{i{\bf k},n}
\apex{2}}{[v\ped{i{\bf k},n}+v\ped{N}]\apex{2}}\delta(E\ped{i{\bf
k}})d\apex{3}k$.\\

$\sigma\ped{i{\bf k}n}(E)$ is the BTK conductance of the $i$-th band
expressed in terms of the quantities {\small $N\apex{q}\ped{i{\bf
k}}(E)=E/\sqrt{E\apex{2}-\Delta\ped{i{\bf k}}\apex{2}}$
}{\normalsize} and {\small $N\apex{p}\ped{i{\bf
k}}(E)=\Delta\ped{i{\bf k}}/\sqrt{E\apex{2}-\Delta\ped{i{\bf
k}}\apex{2}}$ }{\normalsize} (whose real parts are the quasiparticle
and the pair density of states in the same band, respectively) and
of $Z\ped{n}$ values. $\Delta\ped{i{\bf k}}$ is the gap value for
band $i$-th at point ${\bf k}$ over the FS, recently calculated from
first principles \cite{Sanna}. The values of $Z\ped{n}$ used in the
calculation are taken similar to those of the curves shown in
Fig.~\ref{Fig:CaC6_fig2}, i.e. $Z\ped{ab}=$ 0.75 and $Z\ped{c}=$ 1.
The explicit expression of $\sigma\ped{i{\bf k}n}(E)$ can be found
in Ref. \onlinecite{Tanaka}.

\begin{figure}[t]
\begin{center}
\includegraphics[keepaspectratio, width=1.0\columnwidth]{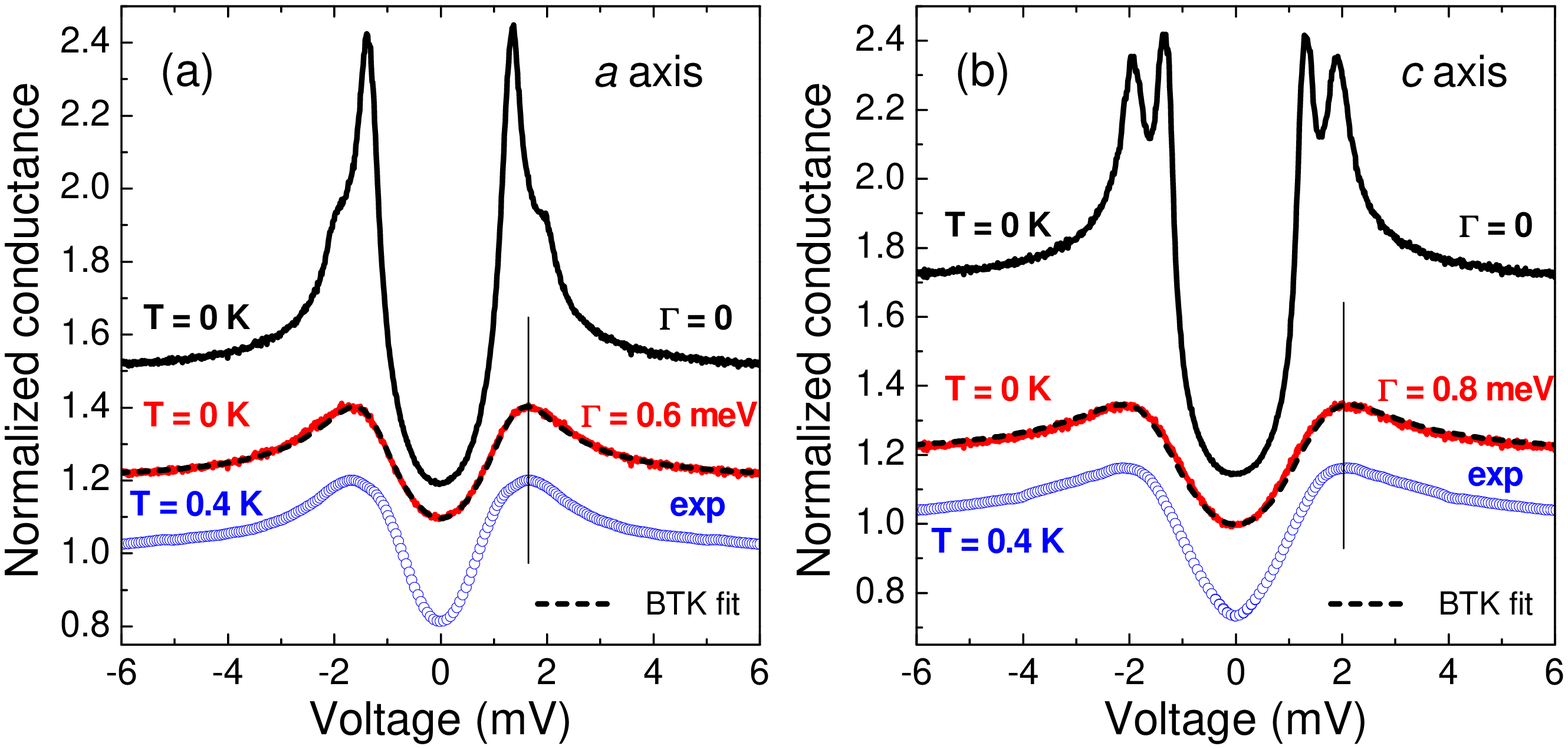}
\end{center}
\vspace{-5mm} \caption{\small{(Color online) Theoretical AR
conductances calculated at $T=$ 0 by Eq. (1). (a) Current injected
along the \emph{a} axis with $Z=$ 0.75 and $\Gamma=$ 0 (black) and
$\Gamma=$ 0.6 (red); (b) current injected along the \emph{c} axis
with $Z=$ 1 and $\Gamma=$ 0 (black) and $\Gamma=$ 0.8 (red).
Experimental curves at 400 mK are shown for comparison (blue
circles).}} \label{Fig:CaC6_fig4} \vspace{-4mm}
\end{figure}

The results of the calculations given by Eq.~(1) are shown at $T=$ 0
K and for \emph{a}- and \emph{c}-axis directions (top curves in
Fig.~\ref{Fig:CaC6_fig4} (a) and (b)). The conductance calculated
along the \emph{b} direction is almost identical to the one in
\emph{a} direction. At $T=$ 0 the topology of the CaC$_6$ FS and the
calculated anisotropy of the $\pi$ and IL gaps result in a sizeable
anisotropy of the AR conductance. In \emph{ab} direction, it
exhibits a sharp peak (related to the $\pi$ gap) at about 1.38 meV
and a broad shoulder (mainly related to the IL gap) at about 1.9
meV. In \emph{c} direction, as expected from the shape of the FS,
the role of the IL (Ca) gap becomes more important and the
conductance shows two distinct peaks with almost the same height.
However, for $\Gamma \neq$ 0 even at $T =$ 0 K these anisotropic
features are rapidly smeared out. The middle curves of
Fig.~\ref{Fig:CaC6_fig4}(a) and (b) show the effect on the
theoretical conductances of a broadening similar to that observed at
very low $T$ in the experimental curves of Fig.~\ref{Fig:CaC6_fig2}.
The conductances become similar to single-gap ones and can be
perfectly fitted by single-gap 3D BTK curves, as shown in
Fig.~\ref{Fig:CaC6_fig4} (black dash lines). The use of more complex
fitting models (anisotropic or two-band BTK) that we tested on our
data does not improve the fit substantially, as already pointed out
in Ref.~\onlinecite{Bergeal,Kurter}. In Fig.~\ref{Fig:CaC6_fig4} the
experimental \emph{ab}-plane and \emph{c}-axis conductances measured
at 400 mK are included too (circles) in order to show the remarkable
agreement with the theoretical curves for the same level of
broadening. Although this broadening washes out the fine anisotropic
structures of the conductance, a clear sign of the underlying gap
anisotropy is still present since the 3D BTK fit of the theoretical
conductances gives different order parameters in the two directions,
$\Delta=$ 1.5 meV ($\Gamma=$ 0.65 meV, $Z=$ 0.765) for \emph{a}-axis
direction and $\Delta=$ 1.7 meV ($\Gamma=$ 0.92 meV, $Z=$ 1.015) for
\emph{c}-axis one. The \emph{c}-axis value is in perfect agreement
with the experimental results (both single curves at 400 mK and the
peak of the distribution of the 14 different contacts). In the
\emph{ab}-plane case, the experimental $\Delta$ values from the
curves at 400 mK and from the peak of the distribution of
Fig.~\ref{Fig:CaC6_fig3} (a) (ranging from 1.3 meV to 1.44 meV) are
smaller than the value obtained from the fit of the theoretical
conductance. This discrepancy could be ascribed to a possible slight
overestimation of the small $\pi$ gap (and, maybe, an
underestimation of the large IL gap associated with Ca FS) in the
theoretical calculations. This fact appears reasonable if one
considers that the first-principle calculations of Ref. \cite{Sanna}
led to an underestimation of $T\ped{c}$ of about 17 $\%$.

In conclusion, the first directional PCAR measurements in CaC$_6$
carried out also at $T=$ 400 mK both along the \emph{ab}-plane and
the \emph{c}-axis direction give strong and reproducible evidence of
the predicted anisotropic nature of the superconducting gap in this
GIC. New first-principles calculations of the expected anisotropy in
the AR conductance curves fully support this conclusion and indicate
that the actual gap anisotropy in CaC$_6$ could be even slightly
greater than theoretically predicted.

We thank Lilia Boeri, O.V. Dolgov, E.K.U. Gross and I.I. Mazin for
useful discussions. This work was done within the projects: PRIN
(No. 2006021741) and Cybersar (cofunded by MUR under PON). V.A.S.
acknowledges the support of RFBR (Proj. No. 06-02-16490).

\end{document}